\documentclass[9pt,twocolumn,twoside]{opticajnl}

\journal{opticajournal} 

\setboolean{shortarticle}{true}

\title{Boundary-Driven Exceptional Points in Photonic Waveguide Lattices}

\author[1,2,*]{Stefano Longhi}

\affil[1]{Dipartimento di Fisica, Politecnico di Milano, Piazza L. da Vinci 32, I-20133 Milano, Italy}
\affil[2]{IFISC (UIB-CSIC), Instituto de Fisica Interdisciplinar y Sistemas Complejos - Palma de Mallorca, Spain}

\affil[*]{stefano.longhi@polimi.it}

\begin{abstract}
We predict and analyze boundary-driven exceptional points in semi-infinite Hermitian photonic waveguide lattices with a side-coupled defect. The exceptional points arise from coherent reflections at the lattice termination, which induce strong memory effects in the defect dynamics. Using an exact analytic approach, we derive the defect's non-Markovian memory kernel, revealing the trajectories and coalescence conditions of the resonances, which can be precisely tuned by the defect position and the coupling strength. Our results provide a simple and experimentally accessible platform for exploring memory-enabled non-Hermitian physics in Hermitian photonic lattices.
\end{abstract}

\setboolean{displaycopyright}{false} 

\begin{document}

\maketitle

{\em Introduction.} 
Exceptional points (EPs) -- spectral degeneracies where both eigenvalues and eigenvectors of a non-Hermitian Hamiltonian coalesce \cite{EP1,EP2,EP3} -- have attracted widespread attention in photonics for their unusual wave dynamics, topology, and enhanced sensitivity (see e.g. \cite{EP4,EP5,EP6,EP7,EP8,EP9,EP10,EP11,EP12,EP13,EP14,EP15,EP15b} and references therein). In most optical implementations, EPs are typically engineered in coupled waveguides or resonator systems by incorporating optical gain and/or loss into an otherwise Hermitian (i.e., power-conserving) structure \cite{
EP5,EP7,EP8,EP9,EP10,EP11}. In this setting, EPs are spectral singularities of non-Hermitian Hamiltonians, often associated to parity-time symmetry-breaking \cite{EP5,EP7,EP10,EP11}.
Interestingly, EPs can also arise in strictly conservative systems governed by unitary dynamics \cite{HEP1,HEP2,HEP3,HEP4,HEP6,HEP7,HEP8,HEP9,HEP10,HEP11,HEP5}.
Such EPs emerge when one focuses on the dynamics of a subsystem of a larger Hermitian system \cite{HEP1,HEP2,HEP3,HEP4}, or equivalently, through dilation approaches that embed an effective non-Hermitian model into an enlarged Hilbert space \cite{HEP6,HEP7,HEP8,HEP9,HEP10,HEP11}.
In the former scenario, EPs usually manifest as the coalescence of resonances -- i.e., the merging of poles of the subsystem Green's function located on the second Riemann sheet \cite{HEP1,HEP2,HEP3,Longhi2025}. A typical case is provided by a set of few waveguides or resonators coupled to a large-dimensional photonic network that acts as a bath \cite{W3,W7}.
For a single resonator or waveguide, the only route to such resonance coalescences is through {\em memory effects}, i.e., non-Markovian dynamics induced by a structured reservoir. Memory-driven EPs are being explored in quantum and wave systems as a novel route to spectral singularities \cite{NM1,NM2,NM3,NM4,NM5,NM6}; however few realizations have so far been reported in photonics using delay-coupled semiconductor lasers \cite{NM3,NM4}.

Side-coupled defect waveguides in infinite or semi-infinite photonic lattices constitute a paradigmatic realization of the Fano-Anderson model \cite{W0,W0b,W1,W2}, where resonance trapping, bound states in the continuum, dark states and Fano interference phenomena have been observed \cite{W3,W7,W0,W0b,W1,W2,W4,W5,W6}. However, the emergence of memory-driven EPs in such settings has been largely overlooked, despite the fact that the lattice termination naturally generates coherent time-delayed feedback.

In this Letter, we reveal boundary-driven EPs in semi-infinite photonic waveguide lattices with a side-coupled defect. An exact analytic treatment based on the Fano-Anderson model provides the defect's memory kernel and the full pole trajectories, showing that coherent reflections at the lattice edge generate and tune EPs through purely Hermitian dynamics. This establishes a simple, experimentally accessible platform where boundary-induced memory effects give rise to non-Markovian EPs without requiring gain, loss, or specially engineered reservoirs.\\

{\em Photonic lattice model and light dynamics.} We consider light propagation in a semi-infinite array of single-mode optical waveguides or resonators with uniform nearest-neighbour coupling rate \(J\). The lattice sites are indexed by \(n=1,2,3,\ldots\), and each waveguide supports a mode with propagation constant \(\omega_{0}\). A defect waveguide with the same propagation constant is side-coupled to lattice site \(n_{0}\ge 1\) with coupling strength \(g\), as schematically shown in Fig.1(a). Within the semiclassical coupled-mode framework, and after moving to a reference frame rotating at propagation constant \(\omega_{0}\), the field amplitudes \(a_{n}(z)\) in the lattice and \(b(z)\) in the defect are governed by the power-conserving system (see e.g. \cite{W2})
\begin{equation}
i\frac{d a_{n}}{dz} = -J(a_{n+1}+a_{n-1}), \qquad n\ge 1,
\label{eq:lattice-wannier}
\end{equation}
\begin{equation}
i\frac{d b}{dz} = -g\,a_{n_{0}},
\label{eq:defect-wannier}
\end{equation}
with Hermitian tight-binding Hamiltonian $H$
and with the boundary condition \(a_{0}(z)=0\) enforcing the semi-infinite termination. We assume that initially the excitation resides in the defect waveguide, i.e., \(b(0)=1\) and \(a_{n}(0)=0\) for all \(n\ge 1\).
 \begin{figure}[h]
  \centering
  \includegraphics[width=0.48\textwidth]{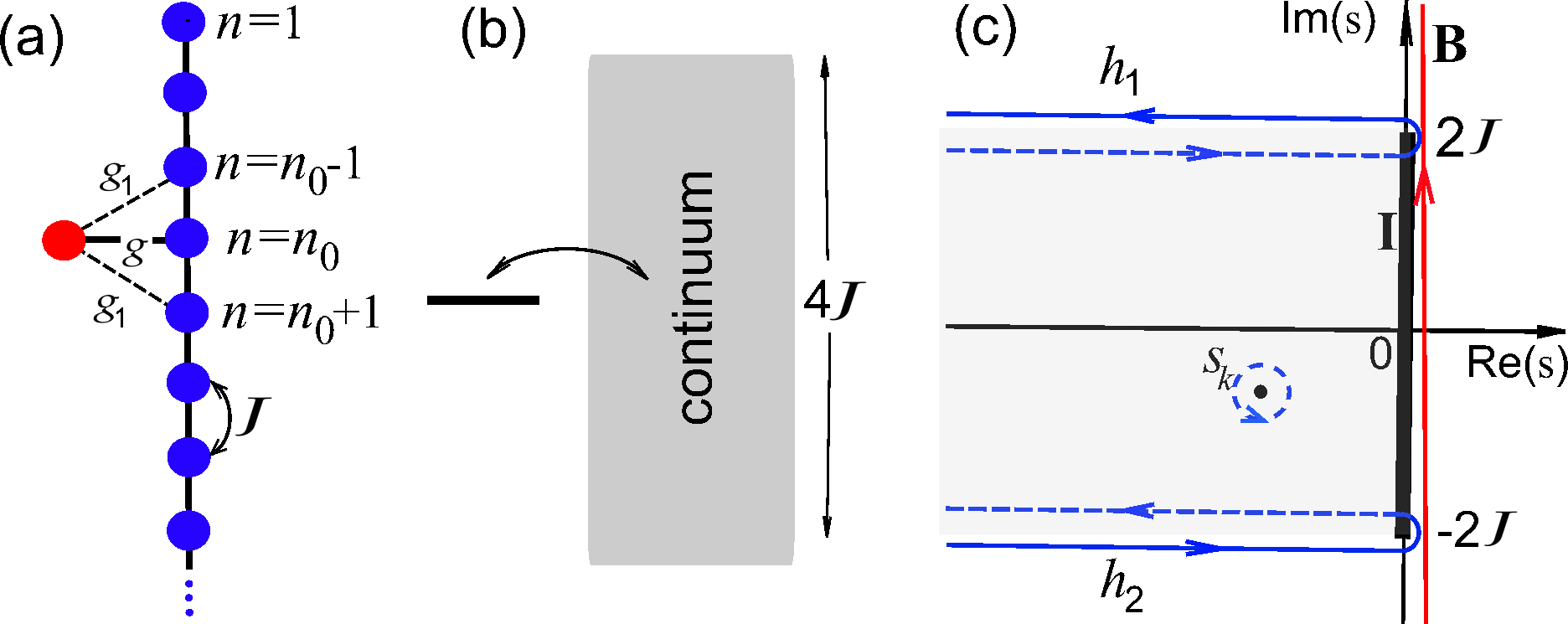}
   \caption{ \small  (a) Schematic of a semi-infinite waveguide lattice with a side-coupled defect waveguide, and (b) corresponding single-level Fano Anderson model. (c) Contour paths in the complex $s$-plane for ~\eqref{bsolution}. The solid segment $\mathbf{I}$ along ${\rm Re}(s)=0$ is the branch cut of the self-energy $\Sigma(s)$. The Bromwich path $\mathbf{B}$ can be deformed into the Hankel paths $h_1$ and $h_2$, and the pole contributions $s_k$ on the second Riemann sheet (resonant states). The shaded region indicates the domain of analytic continuation for $\hat{b}(s)$ on the second Riemann sheet}
 \end{figure}
To eliminate the semi-infinite lattice degrees of freedom, it is convenient to expand the lattice amplitudes into Bloch modes using the sine basis appropriate for a hard-wall boundary (see, e.g. \cite{W1,Pepe}):
\begin{equation}
a_{n}(z)=\sqrt{\frac{2}{\pi}}\int_{0}^{\pi}\! dk\, \sin(kn)\, A(k,z),
\label{eq:sine-transform}
\end{equation}
with \( A(k,z)=\sqrt{2/ \pi} \sum_{n=1}^{\infty} \sin (nk ) a_n(z). \)
Substituting Eq.~(\ref{eq:sine-transform}) into Eq.~(\ref{eq:lattice-wannier}) yields the coupled evolution equation of a single-level Fano-Anderson model
\begin{eqnarray}
i \frac{db}{dz} & = &  - g \sqrt{\frac{2}{\pi}}\, \int_0^{\pi} dk \sin(k n_{0})\, A(k,z) \\
i\frac{\partial A(k,z)}{\partial z} & = & -2J\cos k\, A(k,z)
 - g \sqrt{\frac{2}{\pi}}\, \sin(k n_{0})\, b(z)
\label{eq:Ak-evolution}
\end{eqnarray}
describing the decay of a single level into a tight-binding continuum with a colored interaction \cite{W1} [Fig.1(b)].
As shown in  the Supplemental document, after elimination the bath degrees of freedom the amplitude $b(z)$ satisfies the integro-differential equation 
\begin{equation}
\frac{db}{dz} = \int_0^z dz' K(z-z') b(z') \label{integroequation}
\end{equation}
with the memory function $K(t)$ given by
\begin{equation}
K(t)=-g^2 J_0(2 J t)+(-1)^{n_0} g^2 J_{2n_0}(2Jt), \label{kernel}
\end{equation}
where $J_n(x)$ denotes the Bessel function of the first kind. Equation~(\ref{kernel}) shows explicitly that the dynamics of the defect waveguide is governed by two distinct processes: an instantaneous decay into the lattice, represented by the term \(J_{0}(2J t)\), and a delayed coherent feedback term, represented by \(J_{2n_{0}}(2J t) \), which originates from the reflection of the excitation at the lattice boundary. The latter term vanishes in the $n_0 \rightarrow \infty$ limit.\\ 
The explicit expression of $b(z)$ can be written in terms of a  Bromwich integral by means of Laplace transform using standard methods (see e.g. \cite{W1,Pepe}). One obtains
\begin{equation}
b(z)=\frac{1}{2 \pi i} \int_{-\infty+i0^+}^{\infty+i0^+}ds \hat{b}(s) e^{sz}=\frac{1}{2 \pi} \int_{-\infty+i0^+}^{\infty+i0^+}ds \frac{e^{sz}}{is-\Sigma(s)}
\label{bsolution}
\end{equation}
where $\hat{b}(s)$ is the Laplace transform of $b(z)$ and
\begin{eqnarray}
\Sigma(s) & = & \frac{2g^2}{\pi} \int_0^{\pi} dk \frac{\sin^2(kn_0)}{is+2J \cos k} \\
& = & -i \frac{g^2}{\sqrt{s^2+4J^2}} \left[1-(-1)^{n_0} \left( \frac{\sqrt{s^2+4J^2}-s}{2J}   \right)^{2n_0}    \right] \nonumber \label{selfenergy}
\end{eqnarray}
is the self-energy. The Laplace transform $\hat{b}(s)$ shows a branch cut on the imaginary axis, namely on the segment I defined by $s \in (-2iJ,2iJ)$, corresponding to the continuous spectrum of $H$. Possible poles of $\hat{b}(s)$ can arise, if any, on the imaginary axis ${\rm Re}(s)=0$ and corresponds to bound states (point spectrum of $H$), either inside or outside the continuum. As one can be readily shown, for $n_0$ even $s=0$ is always a pole of $\hat{b}(s)$, corresponding to a compact bound state in the continuum \cite{W1}. Two bound states outside the continuum do exist for $n_0(g/J)^2>2$  \cite{W1}, i.e. they emerge for large $n_0$. Here we focus our attention to the $n_0$ odd case and the weak coupling regime $(g/J)^2 \ll 1$,  such that $b(z)$ decays toward zero.\\
\\
{\em Boundary-driven exceptional points.} To unveil the decay behavior of $b(z)$ and the onset of boundary-driven EPs, it is worth 
deforming the integration Bromwich path B in Eq.(8) as illustrated in Fig.1(c) \cite{Longhi2025,Pepe}.
Deforming the contour requires to cross the branch cut I,  from the right to the left side. Therefore, analytic continuation of $\hat{b}(s)$ on the second Riemann sheet, obtained from Eq.(\ref{bsolution}) by replacing the self-energy $ \Sigma(s)$ with its analytic continuation, should be considered in the shaded region on the left side of the branch cut I. The decay of $b(z)$ has thus two main contributions, one from the poles $s_{k}$ of $\hat{b}(s)$ on the second Riemann sheet (resonances), and  the other one from the
 Hankel path integrals along the contours $h_{1}$, $h_2$ in the first and second Riemann sheets. The latter produces non-exponential decay features at short and long times \cite{Pepe}, while
the poles $s_k$ of $\hat{b}(s)$ on the second Riemann sheet (resonances) govern the multi-exponential relaxation dynamics at intermediate time scale. Each simple pole yields a characteristic exponential  decay contribution $ \sim e^{s_k t}$ with an amplitude determined by the residue of $\hat{b}(s)$ at $s=s_k$.
A non-Markovian EP corresponds to the coalescence of two resonance poles \(s_k\) on the second Riemann sheet. The determination of these poles is discussed in detail in the Supplemental document. In the weak-coupling regime \((g/J)^2 \ll 1\) and $n_0$ not too small, there are two dominant poles, close to the origin $s=0$, which can be approximated as  
\begin{equation}
s_1 \simeq -\gamma +\frac{1}{\tau}\, W_{-1}\!\left(-\tau\gamma\, e^{\gamma\tau}\right),
\qquad
s_2 \simeq -\gamma+ \frac{1}{\tau}\, W_{0}\!\left(-\tau\gamma\, e^{\gamma\tau}\right),
\label{Lambertpoles}
\end{equation}
where 
\begin{equation} 
\tau=n_0/J \; , \; \; \gamma=g^2/(2J)
\end{equation}
and \(W_n(z)\) denotes the \(n\)-th branch of the Lambert \(W\) function~\cite{NM4,L1}. Physically, \(\tau\) represents the round-trip propagation time required for light to travel from the defect waveguide to the lattice edge and back, while \(\gamma\) is the intrinsic decay rate in the edgeless limit \(n_0\to\infty\).
Figure~2 illustrates the behavior of the poles \(s_1\) and \(s_2\) as functions of \(g/J\) for a few increasing  odd values of \(n_0\), comparing the exact numerically computed poles with the analytical approximations of ~\eqref{Lambertpoles}. Note that the agreement becomes excellent as $n_0$ is increased. The figure clearly shows the emergence of an EP: two distinct real poles merge and subsequently split into a pair of complex-conjugate poles sharing the same real part (i.e., decay rate) and opposite imaginary parts.
The EP occurs when \(s_1=s_2\). Using \eqref{Lambertpoles}, this condition yields  
 \begin{figure}[h]
  \centering
  \includegraphics[width=0.5\textwidth]{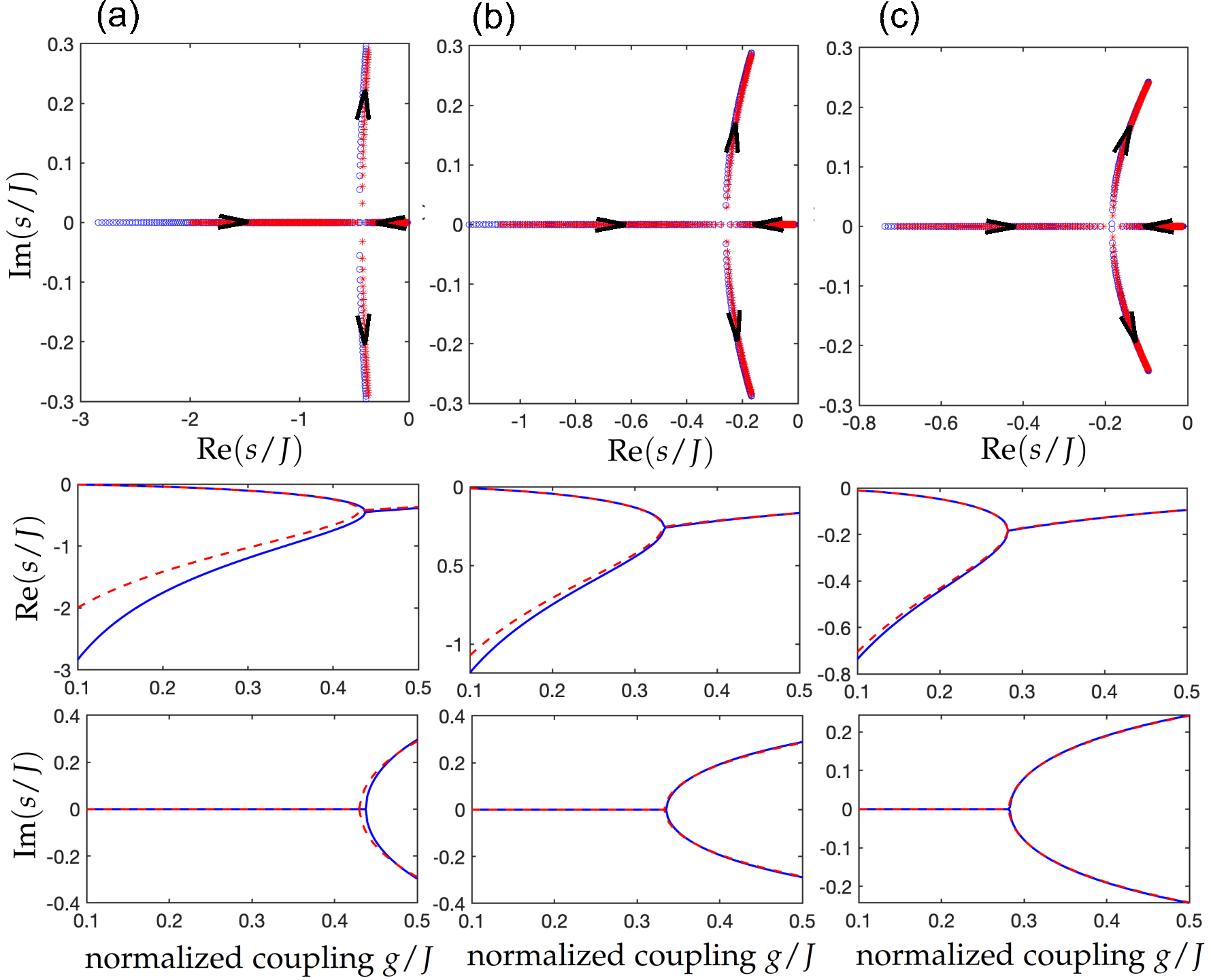}
   \caption{  \small Behavior of the resonance poles $s_1$ and $s_2$ as a function of the normalized coupling constant $g/J$ for increasing odd values of $n_0$: (a) $n_0=3$, (b) $n_0=5$, and (c) $n_0=7$. The solid blue curves (or blue circles) correspond to the exact numerically computed poles, whereas the dashed red curves (or red asterisks) indicate the approximate poles obtained via the Lambert functions. The upper panels show the loci of the two poles in the complex $s$ plane as $g/J$ is varied from $0.1$ to $0.5$ (arrows indicate the direction of increasing $g/J$). The middle and lower panels display the real and imaginary parts of the two poles, respectively, as functions of $g/J$. The boundary-driven non-Markovian exceptional point occurs at the eigenvalue crossing.}
 \end{figure}
\(
\tau\gamma = W_0\!\left( 1/e \right) \simeq 0.27846 ,
\)
which corresponds to
\begin{equation}
(g/J)^2 n_0 \simeq 0.557.
\end{equation}
Crossing this EP leads to a qualitative change in the decay dynamics: the monotonic decay dominated by the slowest pole (with smallest \(|\mathrm{Re}\, s_k|\)) transitions to an oscillatory decay with frequency \(\Omega = 2|\mathrm{Im}\, s_1|\).
Figure~3(a) illustrates this EP behavior through numerically computed decay curves at fixed \(g/J\) and increasing \(n_0\), which drives the system across the EP. The crossing of the EP is clearly manifested as a transition from monotonic decay to damped oscillations of light due to reflections from the lattice edge. Interestingly, the EP condition corresponds to the fastest decay of light from the defect. Indeed, as shown in Fig.~2, at the EP the decay rate -- given by the modulus of the real part of the slowest decaying pole -- attains its maximum value.
 \begin{figure}[h]
  \centering
  \includegraphics[width=0.5\textwidth]{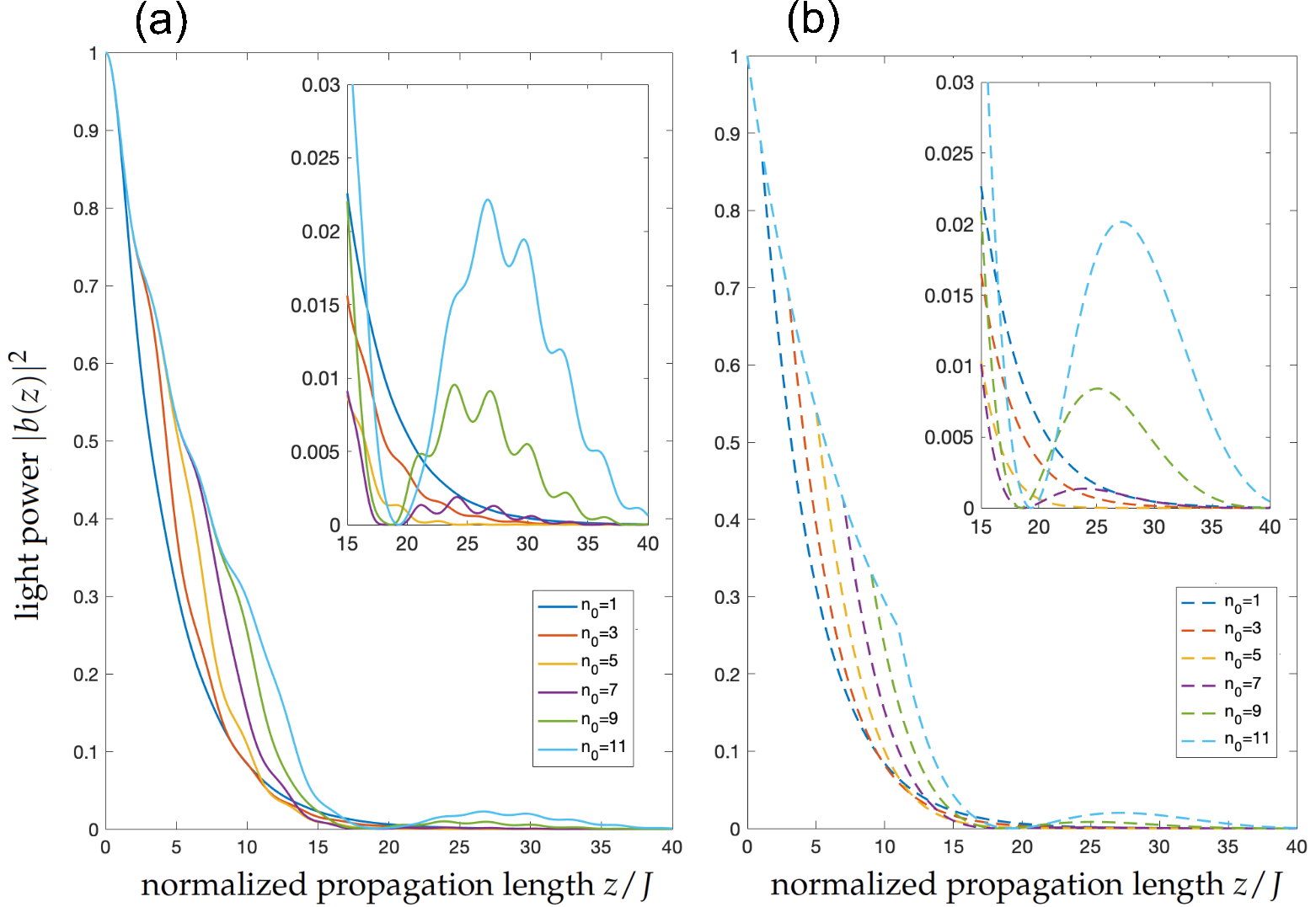}
   \caption{ \small (a) Numerically-computed light intensity decay curves in the defect waveguide for $g/J=0.35$ and for a few increasing values of odd index $n_0$. The EP is crossed as $n_0$ is varied from below to above $n_0=5$. The curves have been obtained by numerically solving the coupled-mode equations (1) and (2) on a lattice comprising 150 waveguides to avoid right edge effects.  The inset shows an enlargement of the decay dynamics in the last stage. As one can clearly seen, crossing the EP as $n_0$ is increased above $n_0=5$ corresponds to the emergence of an oscillatory dynamics, while the fastest decay is obtained at the EP. (b) Same as (a), where the decaying curves are computed using the approximate expression given by Eq.(14) in the text.
}
 \end{figure}
To gain further physical insight into the non-Markovian nature of the EP, it is worth considering the decay dynamics in the framework of the integro-differential equation (\ref{integroequation}). A shown in the Supplemental document, in the weak-coupling regime $(g/J)^2 \ll 1$and for $n_0$ odd  not too small, the equation can be approximated by the differential-delayed equation
\begin{equation}
\frac{db}{dz} \simeq - \gamma \left[ b(z)+  b(z- \tau) \Theta(z-\tau) \right],
\end{equation}
where $\Theta(z)$ is the Heaviside step function. The delayed term in the equation is generated by the lattice edge and is the clear signature of system non-Markovianity \cite{NM3,Tufarelli} .  The solution to this equation can be written as a power series (see e.g. \cite{Tufarelli})
\begin{equation}
b(z) = \sum_{n=0}^{\infty} \frac{\big[- \gamma\, (z-n\tau)\big]^n}{n!}\, e^{-\gamma (z-n\tau)} \, \Theta(z-n\tau), \label{delayedeq}
\end{equation}
where each term in the series represents the contribution to the light amplitude $b(z)$ in the defective waveguide after \(n\) edge bouncing. The decay behavior predicted by the approximate differential-delayed equation (\ref{delayedeq}) is illustrated in Fig.3(b) and reproduces quite well the exact behavior obtained by full numerical simulations. For typical
realistic parameters of waveguide arrays manufactured by femtosecond
laser writing in fused silica \cite{W4,W6,Pepe,decay}, assuming a waveguide
lattice spacing $a \simeq 10 \; \mu$m and a spacing $d \simeq 16 \; \mu$m between the defect waveguide and the lattice
results in coupling constants $J \simeq 4 \; {\rm cm}^{-1}$ and $g/J \simeq 0.35$ at the probing wavelength of 633 nm. The full length of the array in the simulations of Fig.3 is $z \simeq 10 $ cm, which is feasible with current technological platforms. Such light-decay dynamics can be directly monitored using standard fluorescence imaging techniques \cite{W4,Pepe,decay}. Finally, we mention that the transition from monotonous decay to oscillatory damped decay observed in Fig.3 turns out to be quite robust when considering additional long-range coupling terms or weak disorder in the lattice coupling constant $J$, which are unavoidable in realistic experimental settings \cite{Pepe}. 
For example, in Fig.4 we show the light decay dynamics when an additional coupling constant $g_1= \alpha g$ ($ \alpha \ll 1$) between the defect waveguide and the two lattice waveguides at sites $(n_0 \pm 1)$ is included in the analysis, as shown in Fig.1(a).  As one can see, the appearance of the oscillatory decay dynamics as $n_0$ is increased above $n_0=5$ persists even at relatively large values of $\alpha$, demonstrating the robustness of the boundary-driven transition when including additional long-range couplings. The robustness against weak lattice disorder is illustrated in Fig.S2 of the Supplemental document.

  \begin{figure*}
  \centering
  \includegraphics[width=1\textwidth]{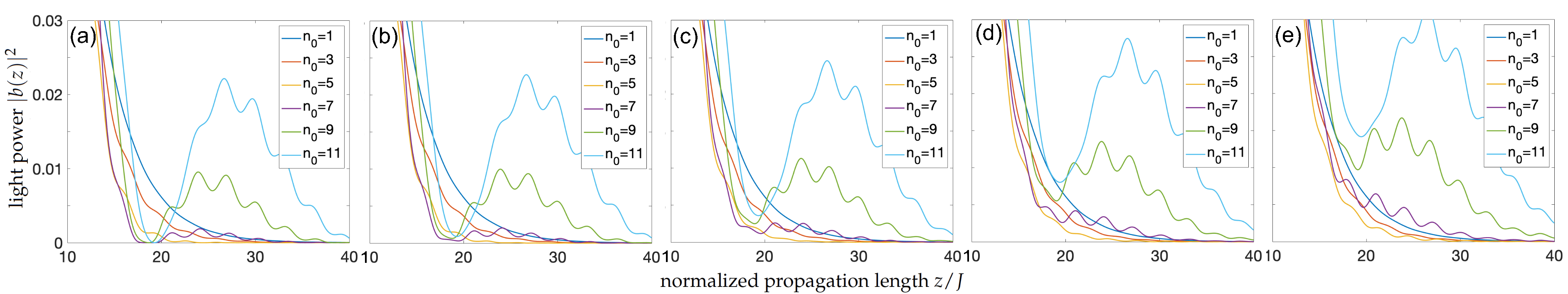}
   \caption{ \small Same as Fig.3(a), but with the additional coupling $g_1= \alpha g$ for (a) $\alpha=0$, (b) $\alpha=0.1$, (c) $ \alpha=0.2$, (d) $\alpha=0.3$, and (e) $\alpha=0.4$. For the sake of visibility, only the last-stage relaxation dynamics is plotted.}
 \end{figure*}

{\em Conclusion.}
We have shown that memory-driven EPs emerge naturally in a fully Hermitian Fano-Anderson model realized with a side-coupled defect in a semi-infinite photonic lattice. An exact treatment of the defect's memory kernel reveals that boundary-induced delay is the key ingredient that drives resonance coalescence on the second Riemann sheet. The resulting boundary-driven EPs are tunable via the defect position and coupling strength. Crossing the EP induces a clear transition from smooth decay to damped oscillations, with the fastest decay occurring at the EP. These results identify coherent boundary feedback as an experimentally accessible mechanism for realizing memory-enabled non-Hermitian singularities in conservative photonic systems. As an application to quantum technologies, the boundary-driven EPs discussed here could be directly relevant to integrated photonic platforms, where the fact that the EP condition corresponds to the fastest relaxation may provide useful design guidelines for relaxation control. Finally, our findings are also expected to be relevant beyond integrated photonics and may be applied in other physical realizations featuring controllable boundary feedback, such as topolectrical circuits \cite{T1,T2}.\\

\noindent
\small
{\bf Disclosures}. The author declares no conflicts of interest.\\
\\
{\bf Data availability}. No data were generated or analyzed in the presented research.\\
\\
{\bf Funding}. Agencia Estatal de Investigacion (MDM-2017-0711).\\
\\
{\bf Supplemental document}. See Supplement 1 for supporting content.

\newpage


 {\bf References with full titles}\\
 \\
 \noindent
 1. T. Kato,
\textit{Perturbation Theory for Linear Operators} (Springer, New York, 1966).\\
2. W.~D.~Heiss,
``The physics of exceptional points,''
J. Phys. A: Math. Theor. \textbf{45}, 444016 (2012).\\
3.M.~V.~Berry,
``Physics of non-Hermitian degeneracies,''
Czech. J. Phys. \textbf{54}, 1039 (2004).\\
4.Y.~N.~Joglekar, C.~Thompson, D.~D.~Scott, and G.~Vemuri,
``Optical waveguide arrays: quantum effects and PT symmetry breaking,''
Eur. Phys. J. Appl. Phys. \textbf{63}, 30001 (2013).\\
5. L.~Feng, R.~El-Ganainy, and L.~Ge,
``Non-Hermitian photonics based on parity-time symmetry,''
Nature Photonics \textbf{11}, 752--762 (2017).\\
6. W.~Chen, S.~Kaya~Ozdemir, G.~Zhao, J.~Wiersig, and L.~Yang,
``Exceptional points enhance sensing in an optical microcavity,''
Nature \textbf{548}, 192--196 (2017).\\
7. R.~El-Ganainy, K.~G.~Makris, M.~Khajavikhan, Z.~H.~Musslimani, S.~Rotter, and D.~N.~Christodoulides,
``Non-Hermitian physics and PT symmetry,''
Nat. Phys. \textbf{14}, 11--19 (2018).\\
8. S.~Longhi,
``Parity-time symmetry meets photonics: A new twist in non-Hermitian optics,''
EPL \textbf{120}, 64001 (2018).\\
9. B.~Midya, H.~Zhao, and L.~Feng,
``Non-Hermitian photonics promises exceptional topology of light,''
Nature Communications \textbf{9}, 2674 (2018).\\
10. M.-A.~Miri and A.~Alu,
``Exceptional points in optics and photonics,''
Science \textbf{363}, eaar7709 (2019).\\
11. S.~K.~Ozdemir, S.~Rotter, F.~Nori, and L.~Yang,
``Parity-time symmetry and exceptional points in photonics,''
Nat. Mater. \textbf{18}, 783 (2019).\\
12. J.~Wiersig,
``Review of exceptional point-based sensors,''
Photonics Research \textbf{8}, 1457--1467 (2020).\\
13. A.~Li, H.~Wei, M.~Cotrufo, W.~Chen, S.~Mann, X.~Ni, B.~Xu, J.~Chen, J.~Wang, S.~Fan, C.~W.~Qiu, A.~Al\`u, and L.~Chen,
``Exceptional points and non-Hermitian photonics at the nanoscale,''
Nat. Nanotechnol. \textbf{18}, 706--720 (2023). \\
14. H.~Meng, Y.~S.~Ang, and C.~H.~Lee,
``Exceptional points in non-Hermitian systems: Applications and recent developments,''
Appl. Phys. Lett. \textbf{124}, 060502 (2024).\\
15. Y.-W.~Lu, W.~Li, and X.-H.~Wang,
``Quantum and classical exceptional points at the nanoscale: Properties and applications,''
ACS Nano \textbf{19}, 19 (2025).\\
16. S. Longhi and L. Feng, "Unidirectional lasing in semiconductor microring lasers at an exceptional point," Photon. Research {\bf 5},  B1 (2017).\\
17. M.~Muller and I.~Rotter,
``Exceptional points in open quantum systems,''
J. Phys. A \textbf{41}, 244018 (2008).\\
18. S.~Garmon, I.~Rotter, N.~Hatano, and H.~Nakamura,
``Analysis technique for exceptional points in open quantum systems and QPT analogy for the appearance of irreversibility,''
Int. J. Theor. Phys. \textbf{51}, 3536 (2012).\\
19. S.~Garmon, G.~Ordonez, and N.~Hatano,
``Anomalous-order exceptional point and non-Markovian Purcell effect at threshold in one-dimensional continuum systems,''
Phys. Rev. Res. \textbf{3}, 033029 (2021).\\
20. T.~Sergeev, A.~Zyablovsky, E.~Andrianov, and Y.~E.~Lozovik,
``Signature of exceptional point phase transition in Hermitian systems,''
Quantum \textbf{7}, 982 (2023).\\
21. U.~Guenther and B.~F.~Samsonov,
``Naimark-dilated PT-symmetric brachistochrone,''
Phys. Rev. Lett. \textbf{101}, 230404  (2008).\\
22. J.-S.~Tang, Y.-T.~Wang, S.~Yu, D.-Y.~He, J.-S.~Xu, B.-H.~Liu, G.~Chen, Y.-N.~Sun, K.~Sun, Y.-J.~Han, C.-F.~Li, and G.-C.~Guo,
``Experimental investigation of the no-signalling principle in parity-time symmetric theory using an open quantum system,''
Nat. Photon. \textbf{10}, 642--646 (2016).\\
23. K.~Kawabata, Y.~Ashida, and M.~Ueda,
``Information retrieval and criticality in parity-time-symmetric systems,''
Phys. Rev. Lett. \textbf{119}, 190401 (2017).\\
24. Y.~Wu, W.~Liu, J.~Geng, X.~Song, X.~Ye, C.-K.~Duan, X.~Rong, and J.~Du,
``Observation of parity-time symmetry breaking in a single-spin system,''
Science \textbf{364}, 878--880 (2019).\\
25. S.~Dogra, A.~A.~Melnikov, and G.~S.~Paraoanu,
``Quantum simulation of parity-time symmetry breaking with a superconducting quantum processor,''
Commun. Phys. \textbf{4}, 26 (2021).\\
26. N.~Maraviglia, P.~Yard, R.~Wakefield, J.~Carolan, C.~Sparrow, L.~Chakhmakhchyan, 
C.~Harrold, T.~Hashimoto, N.~Matsuda, A.K. Harter6, Y.N. Joglekar
and A. Laing, ``Photonic quantum simulations of coupled PT-symmetric Hamiltonians,''
Phys. Rev. Res. \textbf{4}, 013051 (2022).\\
27. L.~Fang, K.~Bai, C.~Guo, T.-R.~Liu, J.-Z.~Li, and M.~Xiao,
``Exceptional features in nonlinear Hermitian systems,''
Phys. Rev. B \textbf{111}, L161102 (2025).\\
28. S.~Longhi,
``Phase transitions and virtual exceptional points in quantum emitters coupled to dissipative baths,''
J. Appl. Phys. \textbf{138}, 180901 (2025).\\
29. S.~Longhi,
``Optical analogue of coherent population trapping via a continuum in optical waveguide arrays,''
J. Mod. Opt. \textbf{56}, 729 (2009).\\
30. S.~Longhi,
``Dark-state photonic entanglement filters,''
Opt. Lett. \textbf{50}, 5101 (2025).\\
31. H.~F.~H.~Cheung, Y.~S.~Patil, and M.~Vengalattore,
``Emergent phases and critical behavior in a non-Markovian open quantum system,''
Phys. Rev. A \textbf{97}, 052116 (2018).\\
32. G.~Mouloudakis and P.~Lambropoulos,
``Coalescence of non-Markovian dissipation, quantum Zeno effect, and non-Hermitian physics in a simple realistic quantum system,''
Phys. Rev. A \textbf{106}, 053709 (2022).\\
33. A.~Wilkey, J.~Suelzer, Y.~N.~Joglekar, and G.~Vemuri,
``Theoretical and experimental characterization of non-Markovian anti-parity-time systems,''
Commun. Phys. \textbf{6}, 308 (2023).\\
34. A.~Wilkey, Y.~N.~Joglekar, and G.~Vemuri,
``Exceptional points in a non-Markovian anti-parity-time symmetric system,''
Photonics \textbf{10}, 1299 (2023).\\
35. J.-D.~Lin, P.-C.~Kuo, N.~Lambert, A.~Miranowicz, F.~Nori, and Y.-N.~Chen,
``Non-Markovian quantum exceptional points,''
Nat. Commun. \textbf{16}, 1289 (2025).\\
36.H.-L.~Zhang, P.-R.~Han, F.~Wu, W.~Ning, Z.-B.~Yang, and S.-B.~Zheng,
``Experimental observation of non-Markovian quantum exceptional points,''
Phys. Rev. Lett. \textbf{135}, 230203 (2025).\\
37. A.~E.~Miroshnichenko and Y.~S.~Kivshar,
``Engineering Fano resonances in discrete arrays,''
Phys. Rev. E \textbf{72}, 056611 (2005).\\
38. A.E. Miroshnichenko, S. Flach, and Y.S. Kivshar,
"Fano resonances in nanoscale structures",
Rev. Mod. Phys. \textbf{82}, 2257 (2010).\\
39. S.~Longhi,
``Bound states in the continuum in a single-level Fano--Anderson model,''
Eur. Phys. J. B \textbf{57}, 45 (2007).\\
40. I.~L.~Garanovich, S.~Longhi, A.~A.~Sukhorukov, and Y.~S.~Kivshar,
``Light propagation and localization in modulated photonic lattices and waveguides,''
Physics Reports \textbf{518}, 1 (2012).\\
41. S.~Weimann, Y.~Xu, R.~Keil, A.~E.~Miroshnichenko, A.~T{\"u}nnermann, S.~Nolte,
A.~A.~Sukhorukov, A.~Szameit, and Y.~S.~Kivshar,
``Compact surface Fano states embedded in the continuum of waveguide arrays,''
{Phys. Rev. Lett.} \textbf{111}, 240403 (2013).\\
42. S.~Longhi,
``Optical analog of population trapping in the continuum: Classical and quantum interference effects,''
Phys. Rev. A \textbf{79}, 023811 (2009).\\
43. A.~Crespi, L.~Sansoni, G.~Della~Valle, A.~Ciamei, R.~Ramponi, F.~Sciarrino, P.~Mataloni, S.~Longhi, and R.~Osellame,
``Particle statistics affects quantum decay and Fano interference,''
Phys. Rev. Lett. \textbf{114}, 090201 (2015).\\
`
44. A. Crespi, F. V. Pepe, P. Facchi, F. Sciarrino, P. Mataloni, H. Nakazato, S. Pascazio, and R. Osellame, "Experimental investigation of quantum decay at short, intermediate, and long times via integrated photonics," {Phys. Rev. Lett.} \textbf{122}, 130401 (2019).\\
45. F. Dreisow, A. Szameit, M. Heinrich, T. Pertsch, S. Nolte, A. T\"unnermann, and S. Longhi, "Decay Control via Discrete-to-Continuum Coupling Modulation in an Optical Waveguide System,"
Phys. Rev. Lett. {\bf 101}, 143602 (2008).\\
46. R. M. Corless, G. H. Gonnet, D. E. G. Hare, D. J. Jeffrey, and D. E. Knuth,
``On the Lambert W function,'{Adv. Comput. Math.} \textbf{5}, 329 (1996).\\
47. T.~Tufarelli, F.~Ciccarello, and M.~S.~Kim,
``Dynamics of a qubit in a structured environment,''
Phys. Rev. A \textbf{87}, 013820 (2013).\\
48.  H. Sahin, M.B.A. Jalil, and C.H. Lee, "Topolectrical circuits -- Recent experimental advances and
developments", APL Electron. Dev. {\bf 1}, 021503 (2025).\\
49. D. Zou, T. Chen, H. Meng, Y.S. Ang,  X. Zhang, and C.H. Lee, "Experimental observation of exceptional bound states in a classical circuit network", Sci. Bull. {\bf 69}, 2194 (2024).

 \end{document}